\documentclass[aps,prb,reprint,superscriptaddress]{revtex4-2}

\pdfoutput=1

\usepackage[utf8]{inputenc}
\usepackage{amsmath,amssymb,amsfonts,graphicx,verbatim,hyperref,mathtools,bbold,float}
\usepackage{physics}
\usepackage{soul, color}
\usepackage{algorithm, algorithmic}
\hypersetup{colorlinks=true, linkcolor=blue, citecolor=blue, filecolor=blue, urlcolor=blue}

\soulregister\ref{7}  
\soulregister\cite{7} 

\DeclareMathOperator*{\argmin}{arg\,min}
\renewcommand*{\eqref}[1]{Eq.~(\ref{#1})}

\newcommand*{\figref}[1]{Fig.~\ref{#1}}

\newcommand*{\secref}[1]{Sec.~\ref{#1}}

\newcommand*{\appref}[1]{Appendix~\ref{#1}}

\begin{document}

\title{Temperature Dependence of Energy Transport in the $\mathbb{Z}_3$ Chiral Clock Model}
\author{Yongchan \surname{Yoo}}
\affiliation{Department of Physics and Joint Quantum Institute, University of Maryland, College Park, Maryland 20742, USA}
\author{Brian \surname{Swingle}}
\affiliation{Department of Physics, Brandeis University, Waltham, Massachusetts 02453, USA}

\begin{abstract}
We employ matrix product state simulations to study energy transport within the non-integrable regime of the one-dimensional $\mathbb{Z}_3$ chiral clock model. To induce a non-equilibrium steady state throughout the system, we consider open system dynamics with boundary driving featuring jump operators with adjustable temperature and footprint in the system. Given a steady state, we diagnose the effective local temperature by minimizing the trace distance between the true local state and the local state of a uniform thermal ensemble. Via a scaling analysis, we extract the transport coefficients of the model at relatively high temperatures above both its gapless and gapped low-temperature phases. In the medium-to-high temperature regime we consider, diffusive transport is observed regardless of the low-temperature physics. We calculate the temperature dependence of the energy diffusion constant as a function of model parameters, including in the regime where the model is quantum critical at the low temperature. Notably, even within the gapless regime, an analysis based on power series expansion implies that intermediate-temperature transport can be accessed within a relatively confined setup. Although we are not yet able to reach temperatures where quantum critical scaling would be observed, our approach is able to access the transport properties of the model over a broad range of temperatures and parameters. We conclude by discussing the limitations of our method and potential extensions that could expand its scope, for example, to even lower temperatures.
\end{abstract}

\maketitle

\section{Introduction}\label{sec:intro}

The study of non-equilibrium quantum systems poses a central challenge across various fields of many-body physics, encompassing classic problems in solid-state systems to more recent problems arising in the dynamics of quantum information~\cite{dalessio2016,nahum2017,white2018,parker2019,skinner2019,xu2020,potter2022,fisher2023}. Here we are interested in a classic problem in the physics of transport, namely the determination of energy currents induced by an applied temperature bias. This problem sits on the border of the broad domain of non-equilibrium physics because the current carrying steady state is a non-equilibrium state, but a notion of approximate local equilibrium still prevails in the system at late time. While the investigation of energy transport is a longstanding endeavor, it remains a challenging problem even in the case of locally interacting one-dimensional systems~\cite{giamarchi2003,bertini2021}, especially if one wishes to probe a temperature regime well below the microscopic energy scales. This paper focuses on energy transport in the context of a non-integrable quantum spin chain, with the specific goal as accessing lower temperatures than previously studied.

To introduce our approach, we first recall that for the simpler problem of equilibrium physics, tensor network approaches, especially matrix product state (MPS) techniques, have demonstrated significant efficacy in one dimension~\cite{schollwock2011,orus2014,paeckel2019,cirac2021}. However, when considering dynamics of a non-integrable system, one generically expects entanglement to grow linearly with time. This growth in turn implies an exponential growth of the requisite bond dimension needed to capture the full state. This is a significant computational barrier, especially since the transport physics of interest is a long-time ``hydrodynamic-like'' property of the system. New approaches have been developed to surpass this entanglement obstacle by devising altered dynamical principles that diverge from the microscopic unitary evolution~\cite{haegeman2011,leviatan2017,white2018,rakovszky2022,lerose2022}, and open quantum system methodologies, which employ an explicit external driving force to guide a system towards its non-equilibrium steady state (NESS), have emerged as a promising avenue for investigating transport properties~\cite{prosen2009,znidaric2010,prosen2012,mendozaarenas2015,znidaric2016,reichental2018,znidaric2019,weimer2021}. On physical grounds, directly connecting a system to a reservoir is anticipated to diminish the entanglement within the system, making it plausible to investigate transport phenomena through a low-entanglement simulation by utilizing an open quantum system methodology.

Nevertheless, while the broad outlines of the open system approach are well established, an important open question is how to design reservoirs which are both efficiently implementable and can drive the system to a wide range of temperatures. There are two broad ways to approach this problem. The first approach involves establishing a reservoir configuration of infinite size and then eliminating the reservoir's degrees of freedom through a tracing-out process. However, the resultant master equation exhibits temporal non-locality, and its memory kernel has so far proven overly intricate for practical solvability~\cite{breuer2007}. For instance, the Redfield master equation~\cite{redfield1965}, which is derived through additional approximations, remains challenging to practically solve.

A second more practical strategy involves searching for an evolution equation for the density matrix of the system that can effectively drive the system towards a controllable equilibrium state or non-equilibrium steady state (NESS). In this situation, one uses the Lindblad master equation~\cite{lindblad1976,gorini1976} with specially chosen jump operators which hopefully drive the system to the desired state. If this is true and if the Lindblad equation can be efficiently solved within a space of low bond dimension matrix product states, then one has a practical method to extract transport physics, as well as other observables. In the case of a system exhibiting good thermalization properties, opting for appropriate local Lindblad operators (which represent the influence of reservoirs) exclusively at the edges of the system leaves the bulk dynamics largely unmodified while restricting the entanglement growth to a manageable amount. It has been argued that many systems, including non-interacting and interacting fermions as well as strongly interacting spins, can achieve thermalization under the condition of infinitely large and weakly damped reservoirs~\cite{reichental2018, zanoci2023}. However, achieving reliable thermalization at low temperatures is made difficult by various obstacles, such as mismatches between the bath scales and the system's energy scales and the possible slow approach to the NESS. Hence, designing jump operators that can effectively drive the system to equilibrium remains a challenging task, one that calls for exploration and experimentation.

Here we study energy transport in a $\mathbb{Z}_3$ chiral clock model~\cite{huse1981,ostlund1981,huse1982,huse1983,haldane1983,howes1983,auyang1987} with boundary driving, with a focus on pushing the current methods to their limits. The chiral clock model has been extensively explored from a theoretical standpoint, driven in part by its relevance to a novel experimental setup involving trapped cold atoms~\cite{bernien2017}. Its low-temperature physics involves a symmetry-breaking quantum phase transition~\cite{sachdev2011}, and a prominent aspect of the model is its distinctive property dynamical critical exponent, $z \neq 1$, at criticality~\cite{zhuang2015,samajdar2018,whitsitt2018}. While substantial advancements have been made in exploring the phase transition characteristics of the model through both field-theoretical and numerically-based Density Matrix Renormalization Group (DMRG) approaches~\cite{white1992}, the dynamical aspects are less well studied. Prior results on energy transport include a generalized hydrodynamics framework at a critical integrable point~\cite{mazza2018}, a NESS approach involving tailored Lindblad operators with a constant bath temperature~\cite{nishad2022} and a DMRG approach for the finite temperature thermal conductivity along a line of integrable points~\cite{manna2023}. But the generic non-integrable behavior as a function of temperature has not yet been explored. Moreover, in light of recent efforts to push to low temperatures with open system methods, the chiral clock model is expected to be a challenging case, at least near the critical line owing to the nearly gapless low-energy spectrum. A long term goal, which we do not achieve here, is to probe transport in the quantum critical regime. The model is also challenging because the local Hilbert space dimension is $3$ as compared to $2$ for the spin-1/2 chains in many other studies. Thus the chiral clock model is both interesting and challenging and provides an excellent opportunity to thoroughly assess the capabilities of the tensor-network-based open system approach.

In this study, we explore the finite temperature transport properties of the $\mathbb{Z}_3$ chiral clock model by imposing a temperature gradient across the system, which can be achieved by manipulating the parameters of the bath operators. As observed in typical non-integrable interacting spin-$1/2$ systems, we anticipate that the model will exhibit a NESS featuring approximate local thermal equilibrium and diffusive energy transport. Importantly, we find that while the expanded local state space considerably increases the computational complexity, it remains feasible to apply the approach to systems over a range of temperatures and system sizes to give good estimates of the transport properties in the thermodynamic limit. The system's effective temperature is evaluated using the thermometry method outlined in Ref.~\cite{zanoci2021}, which relies on comparing steady-state local density matrices to their thermal equilibrium counterparts, quantified through a measure of trace distance.

As an initial exploration, we examine the energy transport within the model for several choices of parameters, focusing on relatively high temperatures for the analysis. As shown in Figure~\ref{fig:gap}, the parameters we study include regimes where the low temperature physics is gapped and points where it is quantum critical. At these high temperatures, we expect and indeed observe conventional diffusive transport irrespective of the model's symmetries and low energy physics (gapped or critical). Next, focusing on parameters where the low energy physics is quantum critical, we assess the temperature-dependence of the energy diffusion constant by progressively reducing the bath temperature. The effective temperature exhibits a linear correlation with the bath temperature before eventually reaching a non-zero saturation point, comparable to the model's characteristic energy scale, $J$. As such, reaching sufficiently low temperatures to directly probe the quantum critical physics remains an outstanding challenge in the open system approach. This behavior mirrors our earlier findings obtained through an analysis of smaller system sizes in Ref.~\cite{zanoci2023}. We also note a resemblance between the temperature dependence of the gapless chiral clock model and the chaotic spin-$1/2$ XZ model~\cite{zanoci2021}.

This paper is structured as follows. First, the $\mathbb{Z}_3$ chiral clock model is introduced in \secref{sec:model}. Then, in \secref{sec:methods}, we outline the boundary open system configuration and methods to estimate the effective temperature and the transport coefficients in question. In \secref{sec:results} we present the finite temperature transport characteristics of the system. Finally, we provide analyses of our findings and explore potential extensions in \secref{sec:discussion}.

\section{The $\mathbb{Z}_3$ Chiral Clock Model}\label{sec:model}

\begin{figure}
\begin{center}
\includegraphics[width=\columnwidth]{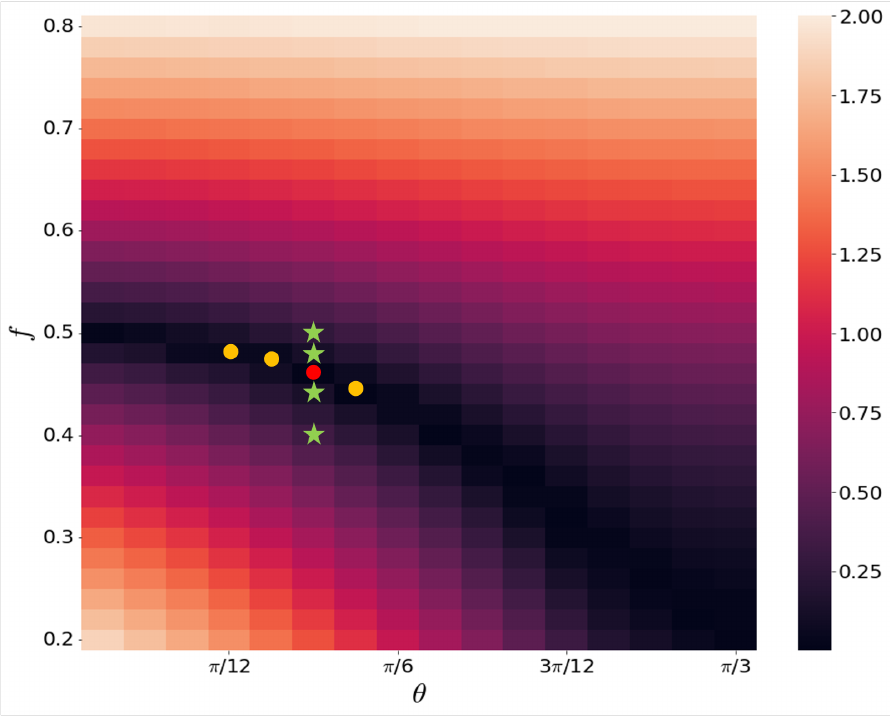}
\caption{The energy gap of the $\mathbb{Z}_3$ chiral clock model for $\phi = 0$. The gap is computed using DMRG on a one-dimensional chain with a length of $L = 200$ and a bond dimension of $\chi = 200$. The depicted black region represents the gapless regime, indicating instances of direct transitions for small $\theta$ values and the intermediate incommensurate regime for larger $\theta$. The yellow and red circles in the middle of the map are the chosen critical points of the energy transport study. The green stars represent selected points to study slightly gapped models. In particular, temperature-dependent transport of the model is considered at the red point.}
\label{fig:gap}
\end{center}
\end{figure}

We consider a one-dimensional $\mathbb{Z}_3$ chiral clock model (CCM) on a chain with open boundary conditions. The Hamiltonian of the CCM for a total chain length $L$ is given by~\cite{zhuang2015,samajdar2018,whitsitt2018}
\begin{equation}\label{eq:model}
    H_{\text{CCM}} = -f e^{-i\phi} \sum_{j=1}^L \tau^\dag -J e^{-i\theta} \sum_{j=1}^{L-1} \sigma_j^\dag \sigma_{j + 1} + \mathrm{h.c.},
\end{equation}
where $\tau_i$ and $\sigma_i$ are the local three-state operators at site $i$. They obey the algebraic relations
\begin{equation}
    \tau^3 = \sigma^3 = \mathbb{1}, \quad
    \sigma \tau = \omega \tau \sigma; \quad
    \omega = e^{2\pi i/3}.
\end{equation}
We choose the explicit matrix representations
\begin{equation}
    \tau = 
    \begin{pmatrix}
        1 & 0 & 0\\
        0 & \omega & 0\\
        0 & 0 & \omega^2
    \end{pmatrix},
    \quad
    \sigma = 
    \begin{pmatrix}
        0 & 1 & 0\\
        1 & 0 & 0\\
        0 & 0 & 1
    \end{pmatrix}
\end{equation}
for $\tau$ and $\sigma$ analogous to the Pauli matrices $\sigma_z$ and $\sigma_x$, respectively, for spin-1/2 systems. From this point of view, the $\mathbb{Z}_3$ chiral clock model can be seen as an extension of the transverse field Ising model, featuring a larger local Hilbert space of dimension $d = 3$. The Hamiltonian presented above contains four parameters: the on-site spin flip strength $f$, the two-site interaction strength $J$, and the two ``chiralities'' $\phi$ and $\theta$.

These many parameters contribute to the model's intricate phase diagram. As implied by its name, the model exhibits a global $\mathbb{Z}_3$ symmetry, which is implemented by the unitary operator $\mathcal{G} = \prod_i \tau_i$. Similar to the behavior of the transverse field Ising model, each coupling strength, $f$ and $J$, defines distinct regions within the phase diagram. Consequently, when $f \gg J$, one can anticipate a disordered phase, while on the opposite side of the phase diagram with $f \ll J$, a $\mathbb{Z}_3$ ordered phase becomes apparent.

The symmetry properties of the model can be further elucidated by the introduction of three operators, charge conjugation $\mathcal{C}$, spatial parity $\mathcal{P}$, and time reversal $\mathcal{T}$. The following symmetry transformation relations are satisfied by these operators~\cite{whitsitt2018}:
\begin{align}
    \mathcal{C}\sigma_i\mathcal{C} = \sigma_i^\dag, &\quad \mathcal{C}\tau_i\mathcal{C} = \tau_i^\dag, \quad \mathcal{C}^2 = \mathbb{1},\\
    \mathcal{P}\sigma_i\mathcal{P} = \sigma_{-i}, &\quad \mathcal{P}\tau_i\mathcal{P} = \tau_{-i}, \quad \mathcal{P}^2 = \mathbb{1},\\
    \mathcal{T}\sigma_i\mathcal{T} = \sigma_i^\dag, &\quad \mathcal{T}\tau_i\mathcal{T} = \tau_i, \quad \mathcal{T}^2 = \mathbb{1}.
\end{align}
The model can have other discrete symmetries depending on the values of parameters $\phi$ and $\theta$ due to the above relationships. When the chiralities are absent ($\phi = \theta = 0$), the model exhibits the presence of all three of these symmetries, leading to the model's reduction to the three-state quantum Potts model~\cite{kedem1993thermodynamics,kedem1993construction}. However, when both $\theta \neq 0$ and $\phi \neq 0$, the discrete spacetime symmetry is solely a composite of $\mathcal{C}\mathcal{P}\mathcal{T}$, with no individual symmetry remaining intact. In contrast, either the $\phi = 0$ or $\theta = 0$ scenario retains separate time-reversal and parity symmetries, wherein the charge conjugation operator $\mathcal{C}$ is coupled with either $\mathcal{P}$ or $\mathcal{T}$, respectively. Notably, the spatial chirality $\theta$ introduces incommensurate floating phases in relation to the periodicity of the underlying lattice~\cite{dai2017}. 

In addition, there is a special parameter curve $f \cos(3\phi) = J \cos(3\theta)$ where the CCM becomes integrable~\cite{baxter2006} and it further exhibits what is known as ``superintegrability'' at $\phi = \theta = \pi / 6$~\cite{albertini1989,mccoy1990}. Integrability offers a broad range of methods to manipulate the model, but transport properties near the integrable point are not fully understood.

For the sake of simplicity, our study focuses exclusively on the CCM with $\phi = 0$ and $f = 1 - J$. This specific choice has been extensively examined in previous literature~\cite{samajdar2018,whitsitt2018}. Notably, we consider several parameter choices for which the low energy physics is quantum critical, consistent with the phase transitions established in Ref.~\cite{samajdar2018}. Because the low energy physics is gapless, we expect that probing low temperatures using the open system approach may be challenging, similar to our previous investigations (Ref. ~\cite{zanoci2023}). We consider a variety of regimes to verify that the low energy physics does not strongly affect the transport physics at moderate to high temperature. The parameters we study are overlayed on a map of the energy gap in Figure~\ref{fig:gap}.

\section{Methods}\label{sec:methods}

\begin{figure}
\begin{center}
\includegraphics[width=\columnwidth]{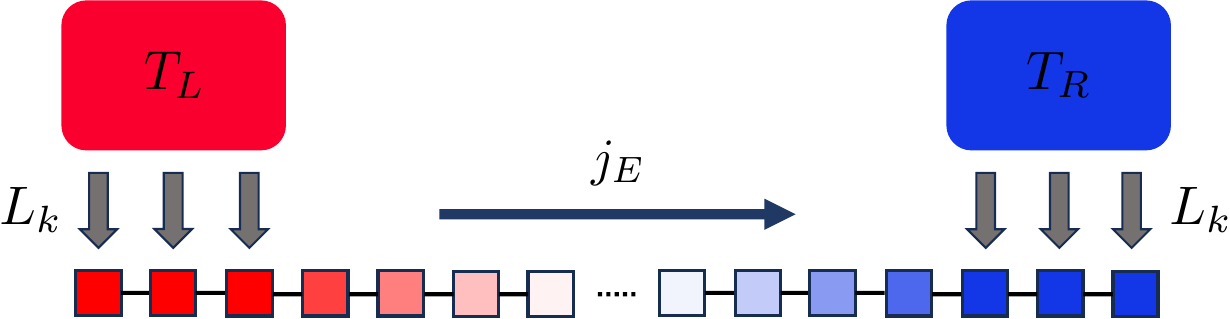}
\caption{Illustrative representation of the boundary-driven transport setup. Thermal baths with temperatures $T_L$ and $T_R$ are produced at both ends of the system using the Lindbladian operator $L_{jk}$. In the depicted scenario, 3-site bath operators are shown, facilitating the establishment of the non-equilibrium steady state. The energy current $j_E$ traversing the system is an outcome of this dynamic arrangement.}
\label{fig:setup}
\end{center}
\end{figure}

\subsection{Tensor Network Simulation Setup}

The non-equilibrium configuration depicted in Fig.~\ref{fig:setup} consists of a one-dimensional chain subjected to Markovian boundary driving, with two bath assemblies connected at its ends. We model the baths using specially chosen jump operators $L_k$ at either end of the chain, with the left and right bath temperatures being separately tunable. This situation is well-characterized by the GKLS Master equation~\cite{lindblad1976,gorini1976}
\begin{equation}
    \frac{d\rho}{dt} = \mathcal{L}(\rho) \equiv -i[H, \rho] + \sum_{jk}\left(L_{jk}\rho L_{jk}^\dagger - \frac{1}{2}\{L_{jk}^\dagger L_{jk}, \rho\}\right), 
    \label{eq:lindblad}
\end{equation}
where $H$ is the Hamiltonian of the system and $L_{jk}$ are Lindblad bath operators, which encode the information of the interaction between the system and the environment, that exclusively operate on the two ends of the one-dimensional chain. The form of these operators is guided by the thermal equilibrium states of the system's Hamiltonian. When the Hamiltonian $H$ is absent, these operators drive the rightmost and leftmost $M$ end sites (bath sites) of the system toward a thermal state $\rho_B$ at temperature $T_B$. A comprehensive description of an arbitrary-size version of $L_{jk}$ is provided in an Appendix of our previous paper~\cite{zanoci2023}.

We start by decomposing the Hamiltonian into three components as $H = H_S + H_B + gH_I$. Here, $H_S$, $H_B$, and $H_I$ represent the sub-Hamiltonians that characterize the system (bulk), bath (both left and right), and their interaction, respectively. In the above decomposition, $g$ is a dimensionless coupling which governs the degree of interaction between the system and the environment. We set $g = 1$ throughout this study to enhance the convergence in time to the NESS. Further elaboration on this matter can be found in \secref{sec:results}.

Consider first the case where $T_L = T_R = T_B$. On general grounds, since the bath sites are only driven to thermal equilibrium when $g=0$ and since the bath is not infinitely large, we do not expect the full system to be in thermal equilibrium at temperature $T_B$ in the steady state. However, as long as the system is still driven to a thermal state and we can determine and tune the system temperature, this setup is still useful. Within our configuration, the temperature gradient is generated by maintaining different temperatures for the left and right bath operators. Specifically, we assign $T_L = 1.2T_B$ to the left and $T_R = 0.8T_B$ to the right bath operator. This deliberate difference in temperature at the two ends is chosen to provide a clear energy profile and energy current. Simultaneously, it ensures a slow variation in the bulk effective temperature along the chain. So long as this variation is small compared to the correlation length in the NESS, we expect a good notion of local thermal equilibrium. Moreover, a small local gradient ensures that the local response, i.e. the induced energy current, is accurately modeled as a linear response. 

We emphasize that $T_B$ will not in fact correspond to the local temperature at the ends chain in the uniform case with $T_L = T_R = T_B$. Similarly, in the biased case, neither $T_L$ nor $T_R$ will correspond to the local temperature at the ends of the chain (see the inset of Figure~\ref{fig:low_transport}(b) for the relationship between $T_B$ and the effective temperature in the middle of the chain). However, as parameters in the jump operators, $T_L$ and $T_R$ do allow us to dial the energy density at either end of the chain, at least within some large range. We separately directly determine the local effective temperature as described just below.

Next, we describe how to find the actual NESS. By considering the superoperator $\mathcal{L}$, which encodes both coherent and dissipative dynamics, the NESS $\rho_{\text{NESS}}$ corresponds to a unique fixed point solution where $d\rho_{\text{NESS}}/ dt = 0$ in~\eqref{eq:lindblad}. It is important to note that while there is often a unique NESS solution, this is not always true~\cite{nigro2019} and the approach to the NESS can be slow even when it is unique. In particular, when jump operators only act on the system's boundary, we can expect many long-lived quasi-steady states. Mathematically, the NESS is equivalent to the limit of the solution of the master equation as time approaches infinity: $\rho_{\text{NESS}} = \lim_{t \rightarrow \infty} \rho(t)$. While some exceptions exist, such as cases involving non-interacting~\cite{prosen2008,prosen2010} and strongly-driven systems~\cite{clark2010,prosen2011,prosen2014,popkov2016,popkov2020}, it remains challenging to directly solve the complete equation for exact NESS solutions in general. However, for the sort of one-dimensional system we are considering, tensor network methods provide a powerful set of tools to represent and evolve the density matrix.

For open quantum system simulations, the vectorization of the density matrix proves highly advantageous for representing the given problem within an expanded Hilbert space~\cite{zwolak2004,weimer2019,landi2022}. This approach involves a superket state, denoted as $\lvert \rho \rangle$, which directly signifies the associated density matrix as a vector within the operator Hilbert space. Simultaneously, two different physical operators $X$ and $Y$ can operate on $\rho$ via $\lvert X \rho Y \rangle = Y^T \otimes X \lvert \rho \rangle$. Within this framework, the Lindbladian operator described in \eqref{eq:lindblad} is transformed into an equivalent Liouvillian superoperator
\begin{align}\label{eq:lindvec}
\begin{split}
    \mathcal{L} &= -i \left(I \otimes H - H^T \otimes I \right) \\
    &\qquad +  \sum_{\nu} \left ( L_{\nu}^{\ast} \otimes L_{\nu} - \frac{1}{2} \left ( I \otimes L_{\nu}^{\dagger} L_{\nu} + L_{\nu}^T L_{\nu}^* \otimes I \right) \right).
\end{split}
\end{align}

Using \eqref{eq:lindvec} as a starting point, the time evolution in the expanded Hilbert space can be realized through the application of the Time Evolving Block Decimation (TEBD) algorithm~\cite{vidal2003,vidal2004} with the superoperator and superket state. To discretize the time evolution operator $e^{\mathcal{L}t}$, we choose the second-order Suzuki-Trotter decomposition~\cite{trotter1959,suzuki1976,paeckel2019} with a time step of $\delta t = 0.05$. The cumulative error arising from these approximations remains sufficiently negligible for the physical parameters employed in our investigation. Given the non-integrable nature of the model, we expect a unique NESS but the convergence time can depend greatly on the initial state. We find it is quite useful to consider an infinite temperature initial state, which is then slowly evolved to the biased finite temperature NESS of interest.  The calculation of the NESS can require a considerable amount of time, depending on the simulation parameters. For the parameters considered here, we found that a simulation time of approximately $t \sim 2000$ generally yields robust convergence. Additional details regarding the simulations are available in \appref{sec:sim_details}.

\subsection{Local Temperature}

Utilizing the tensor network simulation elucidated in the preceding section, we first obtain the NESS of the system for a given set of parameters. Next, given our focus on transport properties as a function of temperature, it becomes imperative to deduce the effective temperature associated with the resulting NESS. As we mentioned, owing to the intricate interplay between the system and its bath, the effective temperature can diverge from the bath temperature $T_B$. One straightforward method to measure the effective temperature involves comparing the final NESS with the Gibbs state at a given temperature, which entails quantifying the dissimilarity between the two states. This approach becomes viable upon the introduction of a notion of local temperature~\cite{hartmann2004,hartmann2005,hartmann2006,garcia2009,kliesch2014}.

We use an approach which allows for an estimation of the effective temperature of the system provided that the system is in local thermal equilibrium within the NESS~\cite{zanoci2021}. The method considers the reduced density matrix $\rho_{\text{NESS}}^A$ of a small local subregion $A$ of the NESS. We also the analogous reduced density matrix $\rho_{\text{thermal}}^A(T)$ obtained by specifying a global thermal state at temperature $T$ for the entire system. We then search for the value of $T$ which minimizes the trace  distance, $K\left(\rho_{\text{NESS}}^A, \rho^A(T)\right)$, between these two states. The trace distance is given by
\begin{equation}
   K\left(\rho_{\text{NESS}}^A, \rho^A(T)\right) = \frac{1}{2}\Tr\left(\sqrt{\left(\rho_{\text{NESS}}^A-\rho^A(T)\right)^2} \right).
   \label{eq:trace_dist}
\end{equation}
and we vary $T$ to minimize this distance measure. This procedure provides a notion of local temperature and tells us how far from local equilibrium the NESS is. Moreover, if the global NESS state happens to be exactly thermal (including the ground state), the procedure will always return the correct global temperature. Physically, we can view this approach as a kind of gradient expansion as discussed in Ref.~\cite{zanoci2021}. The trace distance is a particularly nice comparison tool as it provides an upper bound on the difference in expectation values for any local obserbable~\cite{mendozaarenas2015}.

In our calculations, we chose a pair of adjacent sites as the subsystem $A$, denoted as $(i, i + 1)$. So by dialing $i$ through the chain, we can assign a local temperature to each pair of sites. This choice is not only computationally expedient but also ensures the preservation of a consistent and evenly distributed local temperature across the central region of the system. The temperature at the center of the system, denoted as $T_S$, can be regarded as a representative temperature for the system, derived from the NESS. The expression for $T_S$ is given by:
\begin{equation}
   T_S = \argmin_{T} K\left(\rho_{\text{NESS}}^{\left(\frac{N}{2}, \frac{N}{2}+1\right)}, \rho^{\left(\frac{N}{2}, \frac{N}{2}+1\right)}(T)\right).
   \label{eq:system_temp}
\end{equation}

\subsection{Transport Coefficients}

The NESS serves as a basis for computing the expectation values of any designated local operator. In particular, we can evaluate local expectation values of currents and energy densities and thereby obtain the transport characteristics. For instance, in a scenario where a system exhibits a locally conserved quantity $Q = \sum_i Q_i$, the corresponding local current $j_i$ can be computed utilizing both the continuity equation and Heisenberg's equations of motion:
\begin{equation}\label{eq:discrete_cont}
    \frac{\partial Q_i}{\partial t} = -i \left[ Q_i, H \right] = - (j_i - j_{i + 1}).
\end{equation}
We consider the total energy $E = \sum_i E_i$ as the conserved quantity of the model \eqref{eq:model}, where the local energy operator is represented by the three-site operator
\begin{equation}
    E_i = - f \tau_i -\frac{Je^{i\theta}}{2}\left( \sigma_{i - 1} \sigma_i^\dag + \sigma_i \sigma_{i + 1}^\dag \right) + \mathrm{h.c.},
\end{equation}
with the chiral parameter $\phi$ set to 0. Through a series of algebraic computations employing \eqref{eq:discrete_cont}, one can derive the corresponding energy current operator $j_{E;i}$. This operator can be expressed as a combination of two distinct two-site operators situated at the site $(i, i + 1)$:
\begin{align}
    j_{E;i} &= i\frac{fJe^{i\theta}}{2}\left( j_{E;i}^1 + j_{E;i}^2 \right) + \mathrm{h.c.}\\
    j_{E;i}^1 &= \left( \omega - 1 \right) \sigma_i \left( \tau_i + \tau_{i + 1} \right) \sigma_{i + 1}^\dag\\
    j_{E;i}^2 &= \left( \omega^2 - 1 \right) \sigma_i \left( \tau_i^\dag + \tau_{i + 1}^\dag \right) \sigma_{i + 1}^\dag
\end{align}


We can appeal to Fourier's law as a model of the transport properties. In the context of diffusive energy transport, when an energy bias $\Delta E = \langle E_L \rangle - \langle E_R \rangle$ is maintained across the system, where $\langle E_{L,R}\rangle$ denotes the fixed energy density at the left and right ends respectively, an energy current will be induced in the steady state. Using a continuum approximation with a slowly varying average energy density, the energy current $\langle j_E \rangle(x)$ is related to gradients in the profile of energy density $E(x)$ by
\begin{equation}
    \langle j_E\rangle(x) = - D(E(x)) \frac{d E}{dx}. \label{eq:fourier_loc}
\end{equation}
The current must be independent of $x$ in steady state, so integrating both sides of the above equation gives
\begin{equation}
    L \langle j_E \rangle = - \int_{\langle E_R\rangle }^{\langle E_L\rangle} dE D(E) = - \overline{D} \Delta E .
\end{equation}
In the second equality, we introduced $\overline{D}$ which is the diffusivity averaged over the energy window from $\langle E_R\rangle$ to $\langle E_L\rangle$. Thus we can write
\begin{equation}
\langle j_E \rangle = -\overline{D} \frac{\Delta E}{L},\label{eq:fourier_int}
\end{equation}
with $L$ representing the length of the system. We conclude that the current can be studied as a function of $L$ with $\langle j_E\rangle \propto 1/L$ diagnosing the presence of diffusive transport dynamics. The actual diffusion constant at a given energy can then be obtained from the local form of Fourier's law, \eqref{eq:fourier_loc}.

In practice, the results at the single bond level are slightly noisy and we can get better data by averaging over many sites, as in the integrated Fourier's law, \eqref{eq:fourier_int}. However, $\overline{D}$ will be slightly different from the local $D(E(x))$. To model this, consider the Taylor series $D(E) = D(E_M) + D'(E_M) (E - E_M) + \cdots$ where $E_M = (E_R + E_L)/2$. Integrating, we find $\overline{D} = D(E_M) + \cdots $. Hence, to first order in gradients, $\overline{D}$ is simply $D(E_M)$ and the midpoint energy $E_M$ occurs in the middle of the segment. Thus, using the integrated Fourier's law only entails a small error from the neglect of second order terms. Given the good approximate linearity of the energy profile after discarding edge sites (see Appendix~\ref{sec:sim_details}), and considering our other sources of error, this procedure gives reliable results.



This approach can be expanded to scenarios where the system demonstrates anomalous transport behavior. In such cases, the previously discussed relationship is altered through the introduction of a scaling exponent represented by $\gamma$:
\begin{equation} \label{eq:JvsN}
\langle j_E \rangle = -\overline{D}_\gamma \frac{\Delta E}{L^\gamma}
\end{equation}
In the context of the above equation, we consider a scenario where the only scaling exponent characterizing the transport is $\gamma$. This approach accounts for various forms of transport, including:
(i) ballistic transport ($\gamma = 0$),
(ii) superdiffusive transport $(0 < \gamma < 1)$, and
(iii) subdiffusive transport $(\gamma > 1)$, in addition to the conventional diffusive transport $(\gamma = 1)$.

\section{Results}\label{sec:results}

In this section, we present our findings regarding the finite-temperature transport properties of the $\mathbb{Z}_3$ chiral clock model. The transport coefficients, namely the diffusion constant $D$ and the scaling exponent $\gamma$, are determined from a system size of $L = 48$, which is utilized for all calculations in this section of the paper. Regarding the bath size, we predominantly employ a $2-$site bath configuration for the majority of cases. Furthermore, to tackle more challenging scenarios at lower temperatures, we utilize a $3-$site bath setup and compare it to the $2-$site bath setup. Another critical consideration in our simulations is the bond dimension employed to approximate the resulting NESS. For the chosen parameter ranges, a bond dimension of $\chi = 200$ gives a satisfactory level of convergence, considering the order of magnitude of the trace distance. We provide additional details of convergence in this context in \appref{sec:sim_details}.

\begin{figure*}[t!]
\begin{minipage}{\textwidth}
\includegraphics[width=\columnwidth]{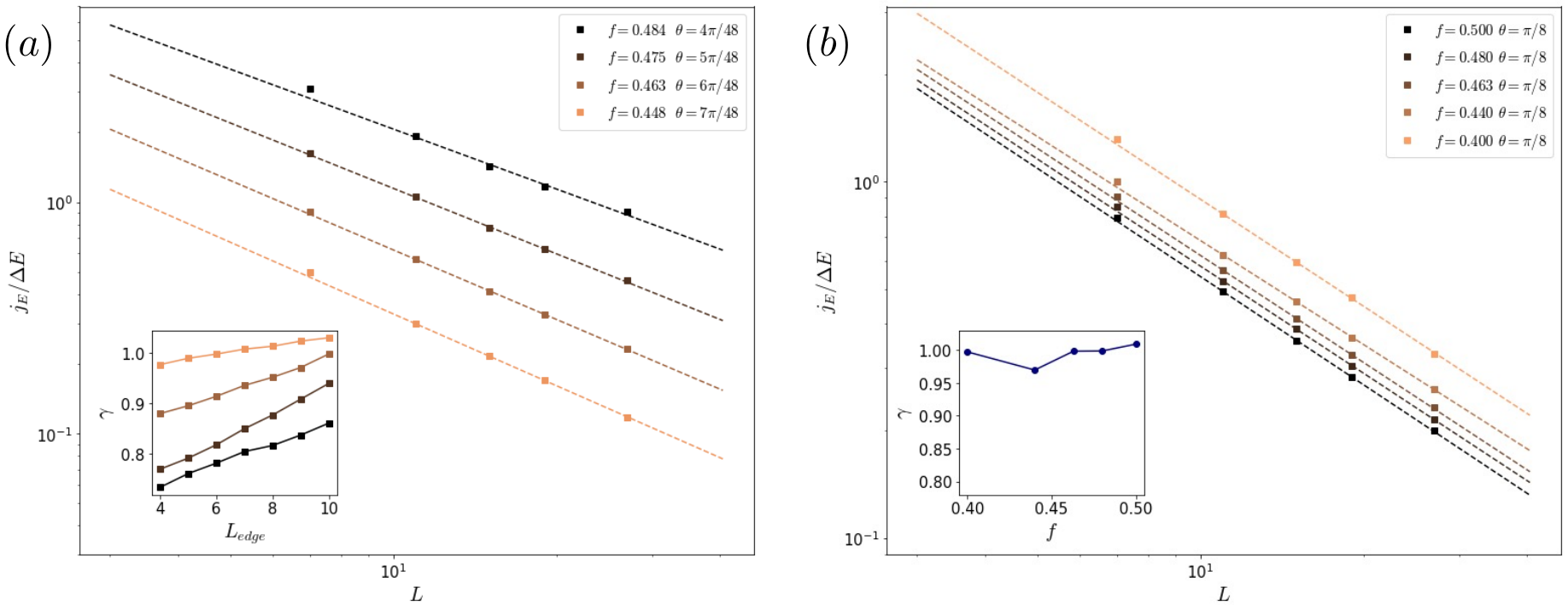}
\end{minipage}
\centering
\caption{The scaled energy current $j_E / \Delta E$ is depicted against the system size $L$ at a high bath temperature of $T_B = 10$. The data is derived from the NESS using selected parameters, encompassing various quantum critical points (a) as well as a slightly gapped regime adjacent to a specific critical point $(f = 0.463, \theta = \pi / 8)$ (b). The dashed lines in the plot correspond to the best-fit results based on the generalized Fourier's law $j_E / \Delta E = -D / L^\gamma$, where $L_{\text{edge}} = 10$ is considered. The top panel also features an inset illustrating the variation of the scaling exponent $\gamma$ with respect to $L_{\text{edge}}$. Meanwhile, the bottom panel includes an inset showcasing the extracted scaling exponent $\gamma$ as a function of $f$.}
\label{fig:high_transport}
\end{figure*}

\subsection{High Temperature Transport Properties}\label{sec:diffusive}

We initiate our investigation by examining the energy transport of the model under conditions of high bath temperature.
In this context, "high bath temperature" refers to a temperature that is finite yet large enough to establish a well-defined effective temperature across the entire system that is large relative to the model's energy scales. In this study, we set the bath temperature to $T_B = 10$ to define this regime of energy transport.
To estimate the system's effective temperature $T_S$, we employ the trace distance calculation outlined in \secref{sec:methods}. In the case of a suitably large system, we determine that the system size does not significantly influence $T_S$. This observation allows us to directly employ Fourier's law to derive the transport coefficient at $T_S$.

The outcomes of the scaled energy current $j_E / \Delta E$ under high-temperature conditions are depicted in Fig.~\ref{fig:high_transport}.
Given that the model becomes non-integrable for non-zero chiralities, it is generally anticipated that the selected points in Fig.~\ref{fig:gap} would exhibit diffusive energy relaxation behavior regardless of the low-energy gap structure.
We start with the results on energy transport in the gapless regime. We choose specific values of $f$ and $\theta$ along the phase transition line, as detailed in Ref.~\cite{samajdar2018}. In the calculation of the expectation values used for estimating transport coefficients, we account for the impact of driving and other boundary effects by excluding a total of $L_{\text{edge}} = 10$ sites at each boundary. Among our chosen points, the influence of finite-size effects from the boundaries is relatively modest for the two cases with larger $\theta$ values.
In these instances, we observe the anticipated diffusive transport behavior, with a scaling exponent $\gamma$ closely aligned with the expected value of 1, accompanied by a small uncertainty of approximately $\pm 0.03$. Conversely, the remaining two points with smaller $\theta$ values exhibit signs of superdiffusive transport, characterized by scaling exponents around the range of 0.8 to 0.9.

Given our anticipation of normal diffusive transport in cases where $T_S \gg J$ and the expected preservation of symmetrical properties, we proceed to examine how finite-size effects impact the scaling exponent $\gamma$ by systematically altering the number of sites excluded at the boundaries, denoted as $L_{\text{edge}}$.
The inset in Fig.~\ref{fig:high_transport} (a) emphasizes the gradual rise of $\gamma$ as the bulk system size becomes more confined to a smaller number of central sites. While not a precise solution, this qualitative examination indicates that strong finite effects tend to exhibit the superdiffusivity.
In general, we observe that the transport coefficients become challenging to compute as we approach the achiral model ($\theta \rightarrow 0$). Specifically, both the energy gradient $\Delta E$ and the energy current $j_E$ become exceedingly small in this region, preventing the transport coefficients from converging with the chosen simulation parameters.

Next, we delve into the analysis of energy transport within the gapped phases of the model. To explore potential variations in transport properties across the direct phase transition line, we focus on a specific point $(f = 0.463, \theta = \pi / 8)$, along with two additional points sharing the same $\theta$ value from both the disordered and $\mathbb{Z}_3$ ordered phases.
This particular point demonstrates better convergence compared to smaller $\theta$ values and remains sufficiently distant from the intermediate incommensurate phase.
Our investigations confirm the prevalence of diffusive energy transport across all the selected points, as depicted in Fig.~\ref{fig:high_transport} (b). Thus, it appears that the model's transport behavior remains robust, unaffected by the low-temperature physics.

We have also extended our method to investigate scenarios where the gap size is comparable to the energy scale ($\Delta \sim J$).
In such cases, the non-integrable model with a larger gap generally exhibits a shorter convergence time~\cite{znidaric2015}.
Similar to the situation with small $\theta$ values, we observe that the scaling exponent $\gamma$ (data not shown) gradually converges towards the normal diffusive value. However, the increasing trend of $\gamma$ is considerably smaller than 1, indicating that convergence of the NESS is affected by other factors, for example, difficulties estimating the gradient of energy and the possible presence of slowly decaying modes in the Liouvillian.
Furthermore, it is noteworthy that the trace distance $K(\rho_{\text{NESS}}, \rho(T_S))$ is primarily influenced by $T_S$. Our observation suggests that as temperature decreases, the trace distance tends to worsen, presenting a challenge in accurately simulating low-temperature physics. While a relatively lower $K(\rho_{\text{NESS}}, \rho(T_S))$ doesn't necessarily ensure the complete reliability of the NESS, it remains a critical criterion for assessing the accuracy of the calculated transport coefficients, considering the computational complexity involved.

\begin{figure*}[t!]
\begin{minipage}{\textwidth}
\includegraphics[width=\columnwidth]{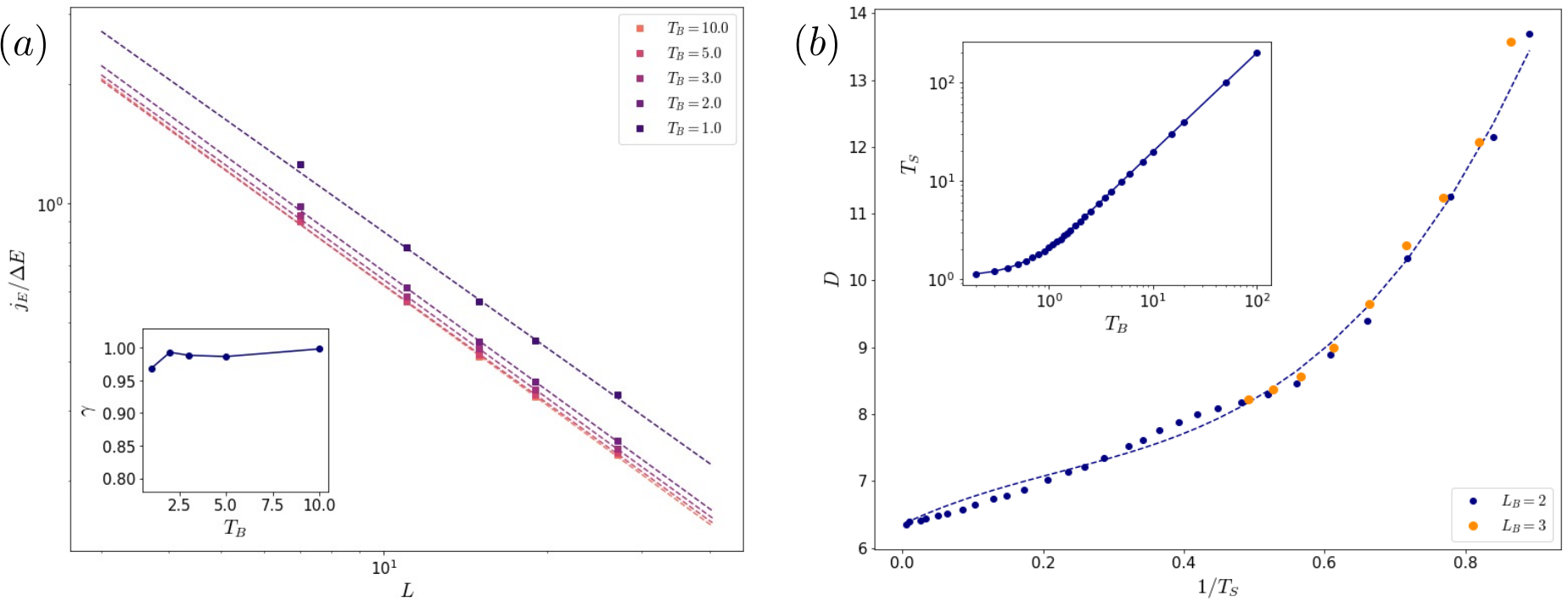}
\end{minipage}
\centering
\caption{(a) The scaled energy current $j_E / \Delta E$ at various bath temperatures is described in relation to the system size $L$, derived from the NESS results for the parameters $(f = 0.463, \theta = \pi / 8)$. The dashed lines illustrate the fitting according to the generalized Fourier's law $j_E / \Delta E = -D / L^\gamma$ with $L_{\text{edge}} = 10$. Inset provides a representation of the scaling exponent $\gamma$ extracted as a function of $f$. (b) The obtained diffusion constant $D$ from the generalized Fourier's law is shown as a function of the reciprocal of the system's temperature $T_S$ for two different bath sizes. The dashed line represents the best-fit power series curve expressed in \eqref{eq:diffusion_high}. The inset plot illustrates the relationship between $T_S$ and $T_B$.}
\label{fig:low_transport}
\end{figure*}

\subsection{Transport Coefficients at Lower Temperatures}\label{sec:transport_coeff}

Now, we proceed to the study of energy transport at various temperatures using the same parameter point as in the slightly gapped regime analysis $(f = 0.463, \theta = \pi / 8)$.
The bath temperature is varied within the range of $0.5 \leq T_B \leq 100$, covering high temperature all the way down to just below the scale of $J$. The scaled energy currents exhibit a proportional relationship with the inverse of the system size within this temperature range, as shown in Fig.~\ref{fig:low_transport} (a).
This result indicates that diffusive energy transport persists even at temperatures comparable to the energy scale $J$.
Here, the trace distance $K$ serves as an indicator of the transport simulation's performance, reflecting the deviation between the NESS expectation values and the thermal state.
For higher temperatures, we observe $K \sim 10^{-3}$, whereas as we approach lower temperatures, the quality of $K$ deteriorates.
As a result, we confine our investigation to temperatures where $K \sim 10^{-1}$ between the steady state and the thermal state, especially for the most challenging simulation scenarios.

Through our thermometry scheme, we observe a linear relationship between the effective system temperature $T_S$ and the bath temperature $T_B$ up to around $T_B \sim 1$. (as seen in the inset of Fig.\ref{fig:low_transport} (b)). 
In contrast, as the temperature decreases to lower limits, we notice that the improvement in $T_S$ becomes non-linear and levels off to a nonzero value, which is on the same order of magnitude as the energy scale of the model, as discussed with a smaller system size in our earlier work (Ref. ~\cite{zanoci2023}). In prior work, we also observed a connection between the resulting $T_S$ and the implementation of the bath.
Specifically, employing a bath with weaker bath-system interactions and larger bath sizes seems to generally enable access to lower temperatures. On the other hand, using a weaker bath can lead to a numerical instability akin to what we observed in the high-temperature transport simulation as $\theta$ approaches zero.
In general, the relaxation time is associated with the spectral gap $\Delta_\mathcal{L}$ of the super-operator.
For our non-integrable model in a boundary-driven setup, a perturbation theory approach suggests that the spectral gap scales as $\Delta_\mathcal{L} \sim \Gamma g^2 / L$ in the limit of $g, \gamma \rightarrow 0$~\cite{cai2013,znidaric2015,medvedyeva2016}.
Consequently, the relaxation time also follows this scaling behavior.
This implies that reducing the strength of bath-system interaction $g$ leads to a notable increase in the time required for the NESS to reliably converge.
However, it turns out that this trade-off provides only a marginal advantage in improving the effective temperature for this specific model.
Despite implementing a larger bath, similar convergence issues persist, and the relaxation time remains largely unaffected by the bath size.
In light of this, we explore the application of the 3-site bath technique in the low-temperature regime, as illustrated in Fig.~\ref{fig:low_transport} (b).
Notably, expanding the bath size primarily offers an enhancement in the trace distance.
Specifically, the trace distance is slightly improved (by approximately $10\%$) for the 3-site bath calculation when compared to the 2-site bath, while the estimated diffusion constants exhibit consistent agreement between the two cases.

Using the approach of estimating local temperature, we acquire temperature-dependent diffusion constants for the model, as depicted in Fig.~\ref{fig:low_transport} (b).
Given that the effective temperature remains significantly higher, making it reasonable to disregard any potential power-law modifications for the zero-temperature limit, we can maintain the assumption of diffusive energy transport within the low-temperature regime.
At high temperatures, the diffusion constant converges to a constant value of approximately $D_\infty \approx 6.3$, exhibiting a notable resemblance to the spin-$1/2$ model outlined in Ref.~\cite{zanoci2021}.
Given the system's gapless nature at the quantum critical point, the energy gap $\Delta$ holds no significance in relation to its transport properties.
Independent of the gap size, a power-series expansion in terms of the inverse temperature is a suitable approach for describing transport in the high-to-intermediate-temperature range, as discussed in Ref. ~\cite{zanoci2021}.
This expansion is expressed as
\begin{equation}\label{eq:diffusion_high}
D = D_\infty \left( 1 + \sum_{a \geq 1}\frac{c_a}{T^a} \right).
\end{equation}
At these temperature levels, a power-series fitting curve with $a = 3$ exhibits excellent agreement with the numerical results, indicating that the energy transport of the model falls within the realm of intermediate temperatures. Given that the model lacks a gap at low energy, one could speculate that a new energy scale needs to be introduced to elucidate this trend within the semi-classical kinetic theory for gapped systems~\cite{damle1998,damle2005,zanoci2021}.

\section{Discussion}\label{sec:discussion}

In this paper, we studied the finite temperature energy transport in the non-integrable $\mathbb{Z}_3$ chiral clock model by utilizing tensor network-based simulations of the open quantum system approach.
Based on the notion of local temperature, the NESS's effective temperature is evaluated using trace distance-based thermometry.
Subsequently, transport characteristics of the model are derived from the NESS using the generalized Fourier's law.
In the high-temperature regime, we observe diffusive energy transport regardless of model parameters so long as we consider the non-integrable regime.
This confirms our expectation that the low-temperature physics does not directly affect the high-temperature transport.
Subsequently, we choose parameters corresponding to a quantum critical point at zero temperature to study the temperature dependence of transport coefficients.
With modest computational resources, we are able to probe transport at lower temperatures comparable to the characteristic energy scale, $J$, of the model.
The use of both $2-$ and $3-$site baths yields consistent outcomes in terms of extracting temperature-dependent diffusion constants at this specific point.
Notably, the resulting diffusion constant at this point in the phase diagram aligns well with a power-series expansion in terms of inverse temperature.

By dedicating significantly larger computational resources, there is potential to delve into energy transport at even lower temperatures. At much lower temperatures, it becomes intriguing to investigate the emergence of power-law behavior in the temperature dependence of the transport coefficients, which could serve as an indication of quantum critical physics in transport. Reaching sufficiently low temperatures in the open system approach remains challenging, but we have taken some steps towards this goal by investigating gains from using a $3-$site bath. Furthermore, exploring the potential correlation between the model's transport properties and its chiralities (and consequently its symmetries) at low temperatures offers another compelling avenue for future exploration.

One of the key messages of our work is that the exploration of these fascinating phenomena can be facilitated through carefully engineered bath configurations. In a previous study (Ref.~\cite{zanoci2023}), a minimum attainable temperature for the gapless chiral clock model was suggested. An interesting avenue for future research could involve developing a comprehensive framework for designing optimized baths and dynamics, aiming to overcome this limitation.
One possible approach to tackle this challenge involves approximating larger baths using the Product Spectrum Ansatz~\cite{martyn2019,sewell2022} for designing dissipators, as larger baths often lead to improved thermalization. 
Another potential avenue is the design of unconventional baths~\cite{larzul2022}, including those based on random energy models or random matrix models.
These bath models tend to exhibit densely populated spectral densities, suggesting that they could lead to effective thermalization if they can be implemented at a sufficiently large scale.
Additionally, exploring similar techniques with other quantum master equations, such as the Redfield equation, could also be considered. 

\begin{acknowledgments}
We thank Cris Zanoci for many valuable discussions. Y.Y. acknowledges support from the U.S. Department of Energy, Office of Science, Office of Advanced Scientific Computing Research, Accelerated Research for Quantum Computing program “FAR-QC”. We also acknowledge computational resources provided by the Zaratan High Performance Computing Cluster at the University of Maryland, College Park.

\end{acknowledgments}

\bibliographystyle{apsrev4-2}
\bibliography{references}

\appendix

\section{Details of Tensor Network Simulations}\label{sec:sim_details}

\begin{figure*}[t!]
\begin{minipage}{\textwidth}
\includegraphics[width=\columnwidth]{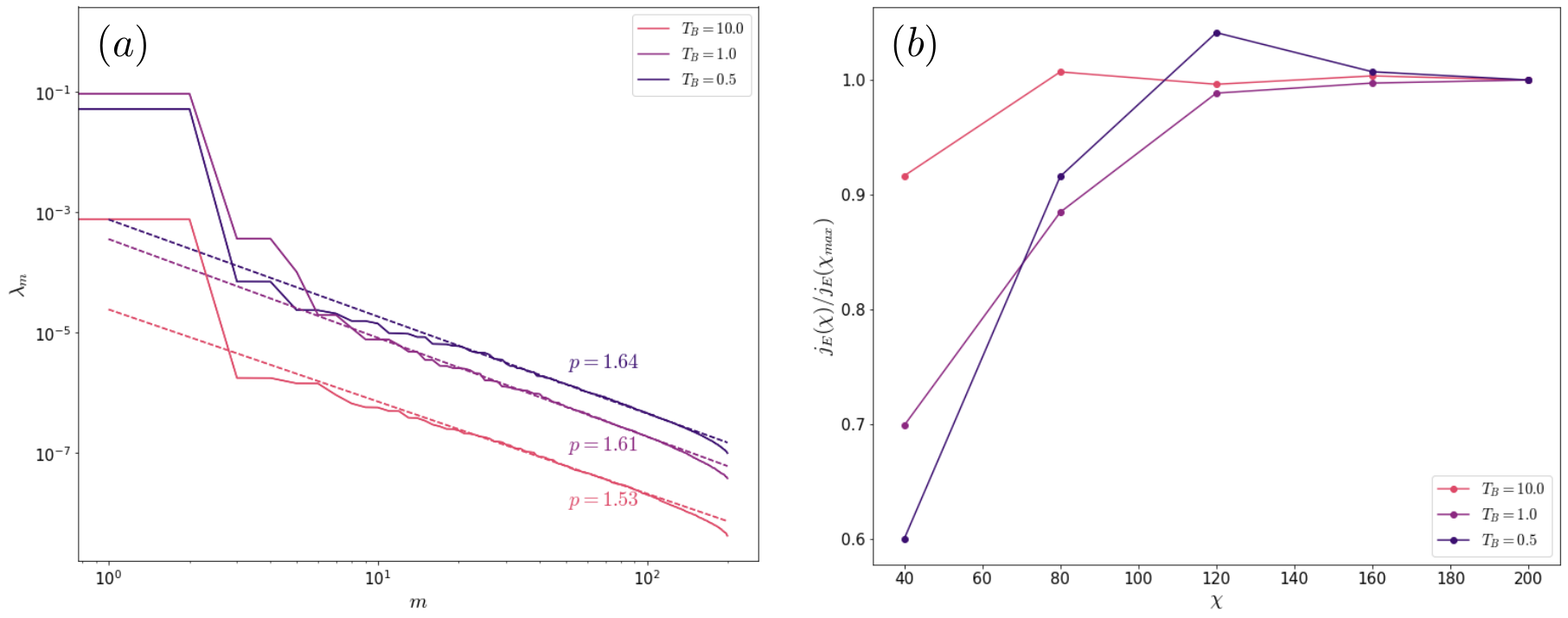}
\end{minipage}
\centering
\caption{Typical convergence of the NESS for $L = 32$, $\chi=200$ at the critical point ($\chi=200, f = 0.463, \theta = \pi / 8)$ at various bath temperatures. (a) Schmidt spectrum at the center of the chain. The best power-law fittings are presented with dashed lines. (b) Convergence of the scaled energy current as a function of bond dimension $\chi$.}
\label{fig:convergence}
\end{figure*}

Our main goal in this text is to investigate how the energy transport of the $\mathbb{Z}_3$ CCM is influenced by temperature variations. In this section, we take a closer look at the details of our tensor network calculations. For our finite temperature tensor network approach, we choose the vectorization approach to depict the density matrix~\cite{zwolak2004} over the purification approach~\cite{verstraete2004}. While this approach doesn't necessarily preserve positivity, it generally offers a more effective way to represent an open quantum system setup compared to alternatives. Using this formalism, we introduce the finite temperature bath into the superoperator $L$, as outlined in Eq.~\ref{eq:lindvec}.

In the context of the boundary driving setup and using the specified superoperators, we attain a non-equilibrium steady state by employing the TEBD algorithm on the vectorized density matrix~\cite{vidal2003, vidal2004,white2004}. The time evolution superoperator is represented as $U = e^{\mathcal{L} \delta t}$ and is discretized using the Suzuki-Trotter decomposition. The $2k-$th order Suzuki-Trotter decomposition of the time evolution operator $U^{(2k)}$ is defined by the following recurrence relation~\cite{suzuki1991},
\begin{align}
    U^{(2)}(\delta t) &= e^{\frac{\delta t}{2}\mathcal{L}_{N - 1}} \cdots e^{\frac{\delta t}{2}\mathcal{L}_1}, \\
    U^{(2k)}(\delta t) &= U^{(2k - 2)}(u_k \delta t)^2 U^{(2k - 2)}((1 - 4u_k) \delta t) U^{(2k - 2)}(u_k \delta t)^2,
\end{align}
where $u_k = 1 / (4 - 4^{1 / (2k - 1)})$.
In our simulations, we opt for a second-order approximation and set the time step to $\delta t = 0.05$. When applying the discretized time evolution operator, we introduce a cumulative error of order $\mathcal{O}(\delta t^{2k + 1})$ into our final state and, consequently, its expectation values. Despite this, we observe that the error remains below $1\%$ within the parameters of our simulation.

An additional significant source of simulation error emerges from truncating the operator Hilbert space dimension of the density matrix. The extent of this truncation error is intricately linked to the operator space entanglement entropy of the NESS. To manage the computational complexity of the simulation, we utilize the standard Schmidt value truncation. In this approach, the Schmidt decomposition of the density matrix into two subsystems, denoted as $A$ and $B$, is expressed as~\cite{prosen2007}:
\begin{equation}
    \rho = \sum_m \sqrt{\lambda_m} \rho_m^A \otimes \rho_m^B.
\end{equation}
Subsequently, the truncation error is evaluated as the sum of truncated Schmidt values, expressed as $\sum_{m > \chi} \lambda_m$. The canonical representation of MPS conveniently grants direct access to the spectrum of Schmidt values~\cite{schollwock2011}. In our simulations, we notice a characteristic asymptotic power-law decay~\cite{bertini2021}, denoted as $\lambda_m \sim m^{-p}$ (refer to Fig. \ref{fig:convergence} (a)), which holds true for all simulation parameters. Hence, we believe that the non-equilibrium state will exhibit an efficient representation in terms of tensor networks even at low temperatures and extended local Hilbert space.

The combination of the cumulative Suzuki-Trotter expansion error and the truncated Hilbert space error introduces an imperfect representation of $\rho_{\text{NESS}}$ as a tensor network. This inherent instability manifests as minor fluctuations in the expectation values, persisting even at later stages of the simulation. To estimate the most probable physical outcome, we take an average of our expectation values over approximately $10^3$ Trotter steps, considering the most challenging simulation in our study.

\begin{figure}
\begin{center}
\includegraphics[width=\columnwidth]{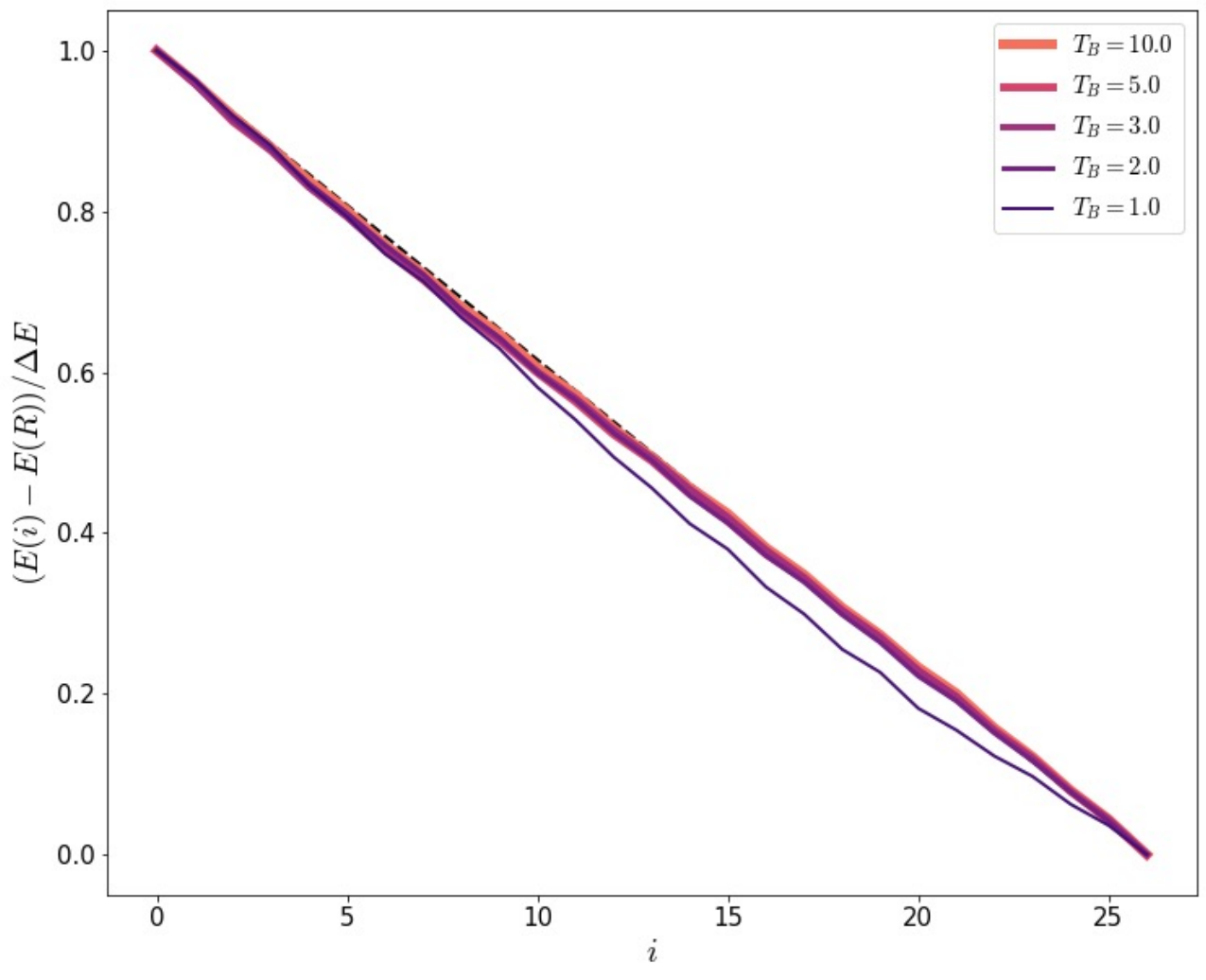}
\caption{The scaled energy profile of the NESS for $L = 48, \chi=200$ at the critical point parameters $(f, \theta) = (0.463, \pi / 8)$ and various bath temperatures for the center region of the system by dropping 10 sites at the each end. The black dashed line represents the exact linear profile.}
\label{fig:energyprofile}
\end{center}
\end{figure}

Typically, opting for the initial state as an infinite temperature state, the dynamics encounters an upsurging entanglement entropy in the early stages, reaching saturation in later times. To ensure a good approximation at each time domain as the density matrix evolves, we employ varying bond dimensions. Initially, we start with a large bond dimension, $\chi = 200$, during the early stages, and then reduce $\chi$ to 81 for intermediate times where the entanglement entropy begins to saturate while the expectation values continue evolving. Subsequently, at late times, we increase the bond dimension back to $\chi = 200$ to fine-tune our solution. We consistently verify the convergence of NESS observables with the bond dimension and find $\chi = 200$ to be sufficient within our range of study. The convergence of the NESS expectation value, as illustrated in Fig. \ref{fig:convergence}, confirms that the NESS energy current improves with a larger $\chi$, aligning with the power-law decay of the Schmidt value spectrum closely tied to the estimated error order~\cite{bertini2021}.

Similar to chaotic spin-$1/2$ systems, our system converges to a unique NESS as a consequence of its dynamics~\cite{nigro2019}. When initializing the system with various states $\rho(0)$, we verify that the resulting state converges to the same NESS, with differences being negligible and attributed to cumulative errors and late-time fluctuations. The uniqueness of this NESS enables us to employ a temperature annealing strategy for lower temperature simulations, in which the relaxation times to reach the NESS are significantly extended. The primary computational workload arises from the gradual evolution during the mid-later stages. However, the early time evolution from an infinite temperature state can be bypassed by initiating the process with a slightly higher temperature NESS to attain the subsequent lower temperature NESS.

\begin{figure}
\begin{center}
\includegraphics[width=\columnwidth]{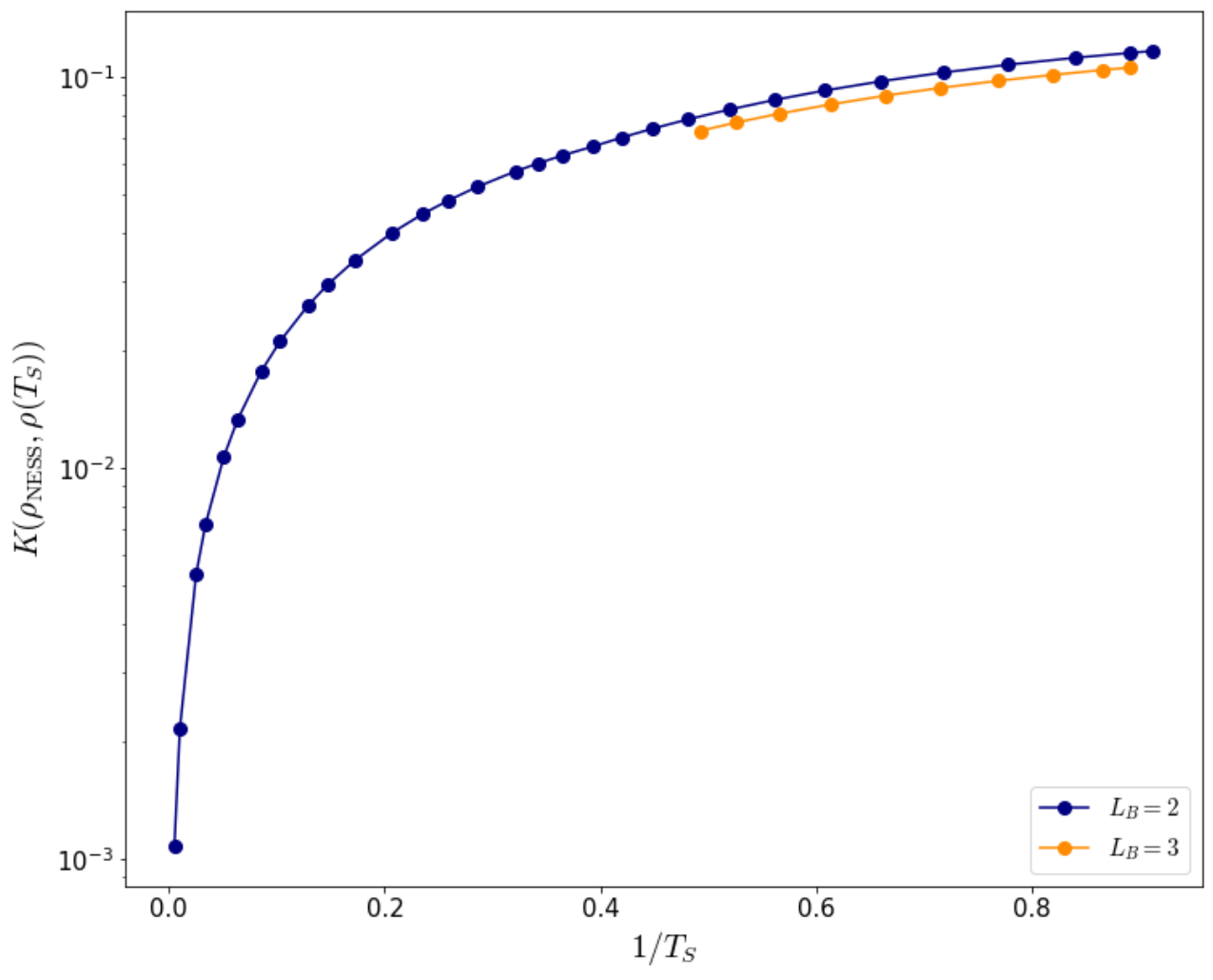}
\caption{The trace distance $K(\rho_{\text{NESS}}, \rho(T_B))$ between the NESS for $L = 48, \chi=200$ and the thermal state for the parameters $(f, \theta) = (0.463, \pi / 8)$ as a function of the inverse of the effective temperature.}
\label{fig:tracedist}
\end{center}
\end{figure}

Obtaining expectation values for local operators with respect to the resulting NESS directly hinges upon the convergence quality of the NESS under the chosen simulation parameters. Given our focus on estimating transport coefficients, one of the most crucial expectation values is the gradient of energy $\Delta E$, which is in the denominator of Fourier's law and typically maintains a small magnitude. Even minor fluctuations in local energy can introduce significant errors in the transport coefficient estimations, limiting our simulation capacity. We carefully choose the central region of the system where the energy gradient closely approximates linearity, as depicted in \figref{fig:energyprofile}, to effectively analyze the bulk transport properties. Thus, we confirm that the analysis presented in the main text remains valid within the scope of our simulations.

In addition to the expectation value of local operators, the trace distance serves as a critical metric for estimating the closeness of the resulting NESS to the local thermal state. As illustrated in \figref{fig:tracedist}, we observe characteristic behavior of the trace distance in the low-temperature simulation discussed in the main text. Notably, the effective local temperature estimated from the trace distance exhibits no significant variance with increasing system size under identical simulation parameters. Consequently, we can leverage Fourier's law phenomenologically to estimate temperature-dependent transport coefficients. However, the trace distance also highlights the challenge inherent in performing accurate low-temperature transport simulations within an open-quantum system framework. While we assert in the main text that a larger external bath marginally improves this situation, the exponential growth in the dimension of the bath operator with its size amplifies computational costs associated with the MPO-MPS multiplication. Furthermore, the relaxation time to reach NESS worsens significantly with larger bath operators, possibly caused by the spectral gap structure of the bath operators. These limitations compel us to restrict the use of bath operators larger than 3 sites for the most stable and reliable method to estimate low-temperature transport properties. Unfortunately, truncating the bond dimension for the thermal bath did not yield a more efficient expression; however, we remain optimistic that advancements in bath engineering could enable low-temperature simulations.

\end{document}